\documentclass[a4paper]{article}

\usepackage{icrc2013}

\def\gtsima{$\; \buildrel > \over \sim \;$}
\def\ltsima{$\; \buildrel < \over \sim \;$}
\def\gsim{\lower.5ex\hbox{\gtsima}}
\def\lsim{\lower.5ex\hbox{\ltsima}}
\def\simleq{\; \raise0.3ex\hbox{$<$\kern-0.75em \raise-1.1ex\hbox{$\sim$}}\; }

\newcommand{\GeV}{{\rm GeV}}

\newcommand{\kpc}{{\rm kpc}}

\newcommand{\km}{{\rm km}}
\newcommand{\muG}{\mu{\rm G}}

\newcommand{\s}{{\rm s}}

\newcommand{\dragon}{{\sffamily DRAGON\ }}

\title{Galactic electron and positron properties from cosmic ray and radio observations}

\shorttitle{Galactic electron and positron properties}

\authors{
D. Grasso $^{1}$,
G. Di Bernardo $^{2}$,
C. Evoli $^{3}$
D. Gaggero $^{4,5}$,
L. Maccione $^{6,7}$
}

\afiliations{
$^1$ Istituto Nazionale di Fisica Nucleare, Sezione di Pisa, Largo B. Pontecorvo, I-56127, Pisa, Italy \\
$^2$ Department of Physics, University of Gothenburg, SE 412 96 Gothenburg, Sweden \\
$^3$ {II.} Institut f\"ur Theoretische Physik, Universit\"at Hamburg, Luruper Chaussee 149, 22761 Hamburg, Germany \\
$^4$ SISSA, Via Bonomea 265, 34136 Trieste, Italy\\
$^5$ Istituto Nazionale di Fisica Nucleare, Sezione di Trieste, Via Valerio, 2, I-34127, Trieste, Italy \\
$^5$ Max-Planck-Institut f\"ur Physik (Werner-Heisenberg-Institut),F\"ohringer Ring 6, D-80805 M\"unchen\\
$^7$ Ludwig-Maximilians-Universit\"at, Arnold Sommerfeld Center, Theresienstra{\ss}e 37, D-80333 M\"unchen.
}

\email{dario.grasso@pi.infn.it}

\abstract{We perform a consistent modeling of cosmic ray electrons, positrons and of the radio emission of the Galaxy.
For the time we reproduce all relevant data sets between 1 GeV and 1 TeV including the recent AMS-02 positron fraction results.  
 We show that below few GeV cosmic ray and radio data require that electron primary spectrum to be drastically suppressed and the propagated spectrum be dominated by secondary particles.  Above 10 GeV an electron + positron extra-component with a hard spectrum is required.
The positron spectrum measured below few GeV is consistently reproduced only within low reacceleration models. We also constrain the scale-height of the cosmic-ray distribution showing that a thin halo ($z_t \lsim 2~\kpc$) is excluded.}


\begin{document}
\maketitle

\section{Introduction}

The latest few years have led to impressive progress in the knowledge of the leptonic component of Galactic Cosmic Rays (CR). 
This was achieved mainly thanks to a set of successful experiments which measured the absolute, combined and relative spectra of electrons and positrons in space or in the high atmosphere. 
Those {\it direct} measurements of CR electrons and positrons (CRE), however, face two major limits: 
1) they are local measurements which may not be representative of the physical conditions in the rest of the Galaxy;
2) below $\sim 20~\GeV$ solar modulation reshape CR spectra respect to their local interstellar spectra (LIS). 
Possible low energy spectral features which may reveal new physics, either related to astrophysical acceleration or to dark matter may, therefore, be hidden. 

Radio observations, by measuring the synchrotron emission produced by electrons and positrons from the entire Galaxy offer a valuable complementary probe of the interstellar spectra and spatial distribution of those CR species.
The interpretation of those measurements requires a proper modeling of their injection, propagation and losses in the Galaxy. 

In this contribution we will summarize the main results obtained by comparing radio and cosmic rays data with the predictions of numerical models of the electron and positron propagation and secondary production in the Galaxy performed with the \dragon package ~\footnote{\url{http://dragon.hepforge.org/}.} \cite{Evoli:2008dv}. 

Respect to our previous result presented in \cite{DiBernardo:2012zu}  here we tune our model against $e^-$ PAMELA data rather than on $e^- + e^+$ spectrum measured by Fermi-LAT.  We also account for the recent AMS-02 positron fraction (PF) results \cite{Aguilar:2013qda} and perform a solar modulation charge dependent treatment \cite{Maccione:2012cu}  which allows to explain the different  low energy results obtained by several experiments.
 
\section{Synchrotron and cosmic ray electron spectra: a consistent model} 

In this section we discuss under which conditions the observed $e^-$ and $e^+$ spectra and their fraction can be modeled consistently with the diffuse Galaxy radio emission between 10 MHz and few GHz.  
We notice that in that frequency interval absorption due to \emph{free-free} scattering is negligible and that
the diffuse radio emission of the Galaxy is almost entirely due to the synchrotron radiation of CR electrons and positrons propagating in the Galactic magnetic field (GMF).  

For the GMF regular component we adopt here a recent model \cite{Pshirkov:2011um}  which is based on a wide and updated compilation of Faraday rotation measurements. Concerning the random component ${\vec B}_{\rm ran}$, we assume it to fill a thick disk with a vertical profile and a scale height $z_t$ from the Galactic plane. 
Differently from previous analysis, where the diffusion coefficient $D$ was assumed to be spatially uniform in the thick disk, we assume $D(z) \propto B_{\rm ran}^{-1}(z)$ as expected from quasi linear CR diffusion theory. Concerning its the rigidity dependence, we assume   
%
$ \displaystyle D(\rho) = D_0 ~\beta^\eta \left(\frac{\rho}{\rho_0}\right)^\delta$
%
with $\beta$ being the particle speed in units of $c$. 

We consider a representative set of models with different choices of the parameter $\delta$, of the scale height $z_t$ and of the Alfv\`en velocity setting the re-acceleration strength,
which we report in Tab.\ref{tab:models}. 
Those models have been tuned to reproduce the observed B/C and proton spectrum measured by PAMELA with a low $\chi^2$. Different models, however, predict quite different secondary electrons and positrons spectra. We will see that this property can actually be used to reject some among the most commonly used propagation setups. 
For what concerns CR leptons,  PAMELA \cite{Adriani:2008zr,Adriani:2010ib} , Fermi-LAT \cite{FermiLAT:2011ab} and AMS-02 \cite{Aguilar:2013qda} PF results require an $e^\pm$ extra-component of the form $J(e^\pm) \propto E^{- \gamma(e^\pm)}~\exp(- E/E_{\rm cut})$, with $E_{\rm cut} \simeq 1~{\rm TeV}$ to be introduced. Above 10 GeV, we tune the background and extra-component spectral indexes to consistently match the $e^-$ spectrum measured PAMELA \cite{Adriani:2011xv} and the PF measured by AMS-02 (which agrees with PAMELA between 10 and 250 GeV).  Below that energy, we set the source spectral index of the $e^-$ background component  requiring  that the predicted synchrotron spectrum agrees with radio observations.  

\begin{table}[tbp]
\centering
  \begin{tabular}{|c|c|c|c|c|c|}
    \hline
    {\bf Model} & $\delta$ & $v_A (\km/\s)$ &  $\eta$ & $\gamma_0(e^-)$ &  $\gamma(e^\pm)$  \\
    \hline
     PD & 0.6 & 0 &  -0.4 & 1.2/2.5 & 1.75  \\
     KRA & 0.5 & 15 & -0.4 & 1.2/2.5  & 1.75  \\
     KOL & 0.33 & 34 & 1 & 1.2/2.5 & 1.75  \\
    \hline
  \end{tabular}
\caption{ \label{tab:CRE_models} The main parameters of the models considered in this work. The reported values of $\gamma_0(e^-)$ refer to $E$ smaller/larger than $4~\GeV$. For each of those setups we considered several value of the scale height $z_t$ which amount to a proper rescaling of $D_0$.}
\label{tab:models}
\end{table}

Similarly to what done in \cite{DiBernardo:2012zu} we use here a 2-dimensional version of \dragon which adopts a CR distribution invariant for rotations about the Galactic disk axis. 
This is well suited to model the CRE propagation below 10 GeV where energy losses can be neglected. 
At larger energies the spiral arm distribution of astrophysical sources cannot be neglected since the energy loss length become comparable, or smaller, than the solar system distance from the closest arms. In \cite{Gaggero:2013rya,Gaggero_ICRC13} this problem was addressed using a 3-dimensional upgrade of the \dragon code.
 Below few hundred GeV and for the purposes of this work, however, this effect just amounts to a softening of the electron spectrum which can be effectively compensated by an hardening of source spectrum respect to that determined with the 2D code.  Noticeably, such a hardening helps reconciling our models with Fermi acceleration theory and radio observation of SNRs.  

For each CRE model we compute the synchrotron emission at different frequencies by performing, for a given position in the sky, a line of sight integral of the synchrotron emissivity as given in Eq. (3.1 and 3.2) of \cite{DiBernardo:2012zu}.
Then we use {\tt HEALPix}~\footnote{\url{http://healpix.jpl.nasa.gov}.} to properly average the resulting flux over the sky regions $40^\circ < l < 340^\circ$, $10^\circ < b < 45^\circ$ and $-45^\circ < b < -10^\circ$where $l$ and $b$ are Galactic longitude and latitude respectively. This is the region where the contamination from point-like and local extended sources is expected to be the smallest.  In this region we compare the simulated spectra with the ones measured by a wide set of radio surveys  at 22, 45, 408, 1420 , 2326 MHz as well as WMAP foregrounds at 23, 33, 41, 61 and 94 GHz as consistently catalogued in \cite{deOliveiraCosta:2008pb}. 

\begin{figure}[t]
  \centering
  \includegraphics[width=0.5\textwidth]{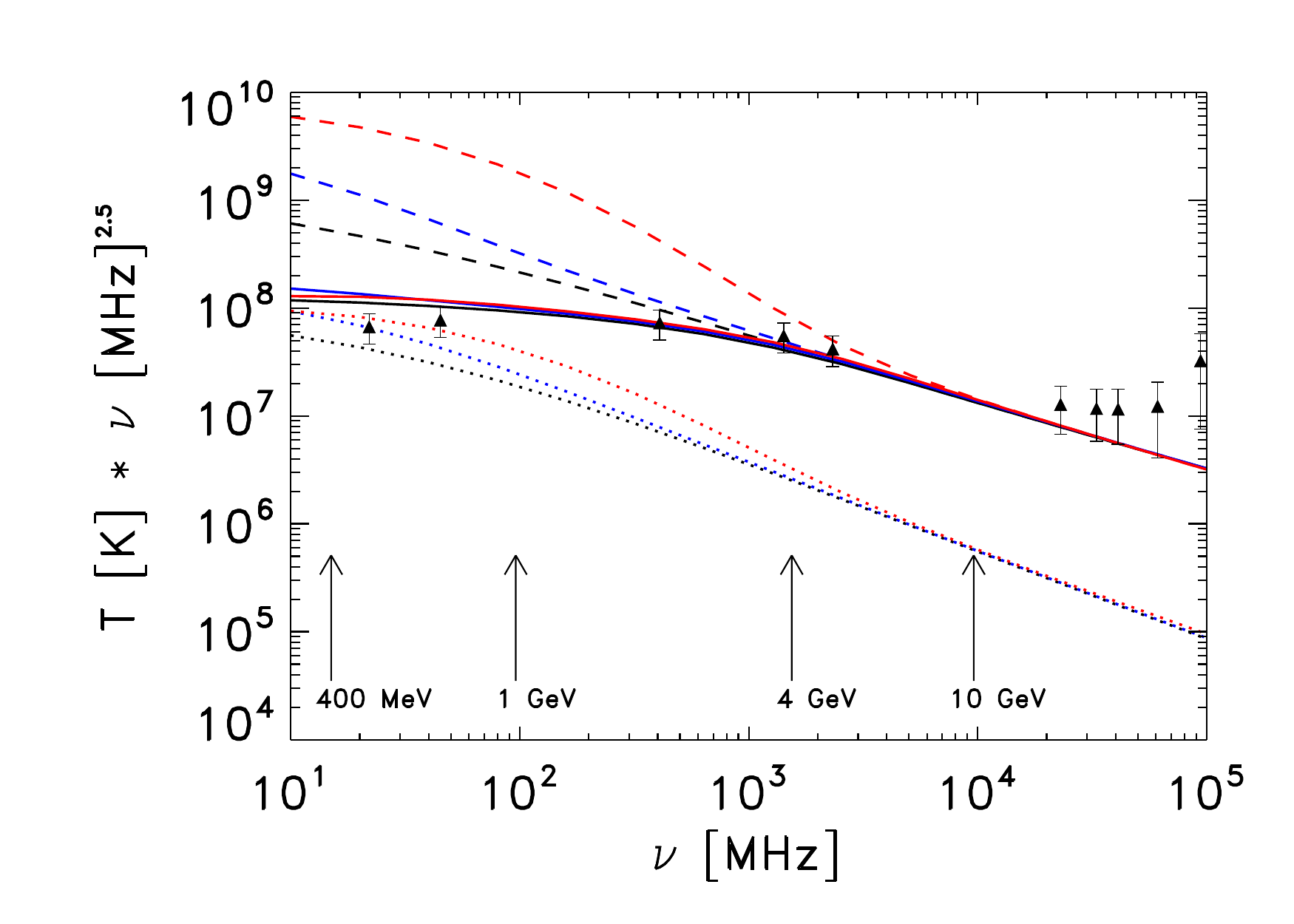}
  \caption{The average synchrotron spectra in the region $40^\circ < l < 340^\circ$, $10^\circ < b < 45^\circ$ computed for the reference propagation setups PD (black lines),  KRA (blue) and KOL (red) defined in table \ref{tab:models} are compared with experimental data derived from \cite{deOliveiraCosta:2008pb}. For each setup we show the spectra obtained with (solid lines) and without (dashed line) the spectral break in the $e^-$ source spectra given in table \ref{tab:CRE_models}. Dotted lines represent the corresponding contribution of secondary $e^-$ and $e^+$.  The random component field strength is tuned to reproduce the spectrum normalization at 408 MHz. The required normalization is $B_{\rm ran}(0) = 7.6~\muG$. The critical synchrotron frequencies, calculated for this value of $B_{\rm ran}(0)$, are reported for a few reference values of the electron energy.   Microwave data above 20 GHz are expected to be contaminated by non-synchrotron emission. Therefore they only provide upper limits to the synchrotron flux and are shown here only as a reference. }
  \label{fig:synchro}
 \end{figure}

\begin{figure}[t]
  \centering
  \includegraphics[width=0.5\textwidth]{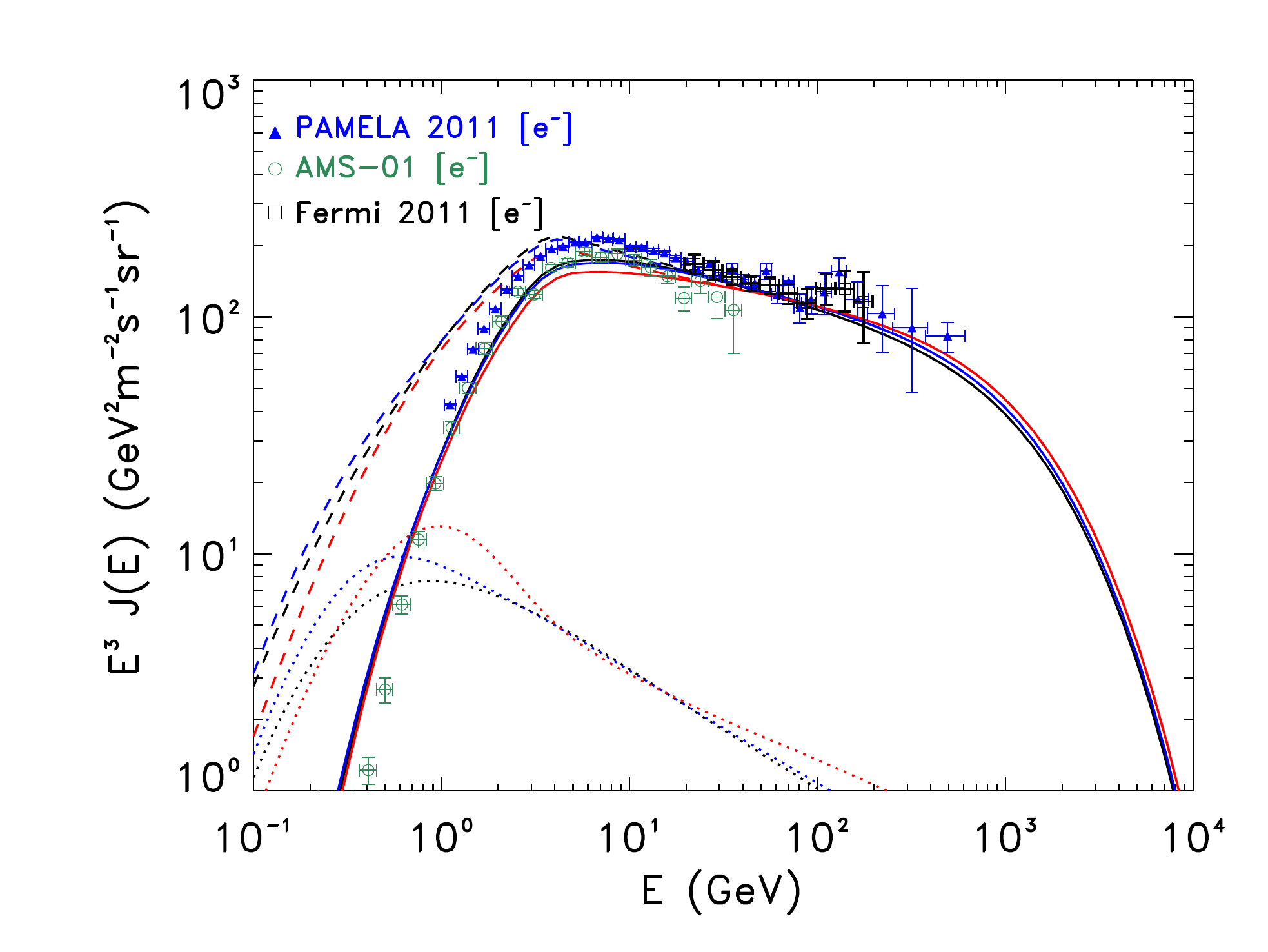}
  \caption{The $e^-$ computed for the reference propagation setup PD (black lines), KRA (blue) and KOL (red) defined by the parameters in table \ref{tab:CRE_models} and $z_t = 4~\kpc$ are shown together with a selection of experimental data sets. 
Continuos and dashed lines represent modulated (AMS-01 time) and interstellar (LIS) spectra respectively. The dotted lines correspond to the secondary contributions to the LIS spectra.}
  \label{fig:electrons}
 \end{figure}

From Fig.\ref{fig:synchro} it is evident that radio data are incompatible with a single power-law electron spectrum.
Rather, we find that $e^-$  broken source spectra characterized by the spectral indexes reported in Tab.\ref{tab:models} provide very good descriptions of radio data. 
A relevant consequence of the break in the $e^-$ spectrum is that below few GeV the $e^-$ flux (see Fig.\ref{fig:electrons}), hence the radio spectrum below $100~{\rm MHz}$ (see dotted lines in Fig.\ref{fig:synchro}), are dominated by secondary particles offering a probe of the interstellar proton spectrum. 
Noticeably, for the low reaccelerating models (PD and KRA) a single power law proton spectrum which agrees with that measured by PAMELA \cite{Adriani:2011cu} below few hundred GeV reproduces very well all data sets. 

We then check if the low energy electron spectra tuned to reproduce radio data are compatible with direct CR measurements, namely with 
the $e^-$ spectrum measured by AMS-01 \cite{Aguilar:2007yf} and PAMELA \cite{Adriani:2011xv} as well as the PF measured by PAMELA \cite{Adriani:2008zr,Adriani:2010ib} and AMS-02 \cite{Aguilar:2013qda}. 
We do that by treating solar modulation as described in \cite{Maccione:2012cu}. This involve three main parameters: the solar magnetic field polarity $A$; the current sheet tilt angle $\alpha$ and the particle mean free path normalization $\lambda$ measured in astronomical units. The values of those parameters during the data-taking periods of the relevant experiments are reported in Tab.\ref{tab:helio_par}.   

\begin{table}[tbp]
\centering
  \begin{tabular}{|c|c|c|c|}
    \hline
    {\bf Experiment} & $A$ & $\alpha$ &  $\lambda_0$~(AU)  \\
    \hline
     AMS-01  & +1 & $10^\circ$  & 0.4   \\
     PAMELA & +1 & $10^\circ$  & 0.4   \\
     AMS-02  &  -1 & $60^\circ$  & 0.15   \\
    \hline
  \end{tabular}
\caption{The heliosphere parameters during the PF measurements of AMS-01, PAMELA and AMS-02 experiments.}
\label{tab:helio_par}
\end{table}

From Fig.\ref{fig:posfrac} we see as the PF and $e^-$ spectrum are consistently reproduced under those condition for the PD and KRA models.
Rather, strong re-acceleration models (KOL model here) produce too much secondaries $e^+$ at low energy and are clearly incompatible with AMS-02 data. 

 \begin{figure}[t]
  \centering
  \includegraphics[width=0.5\textwidth]{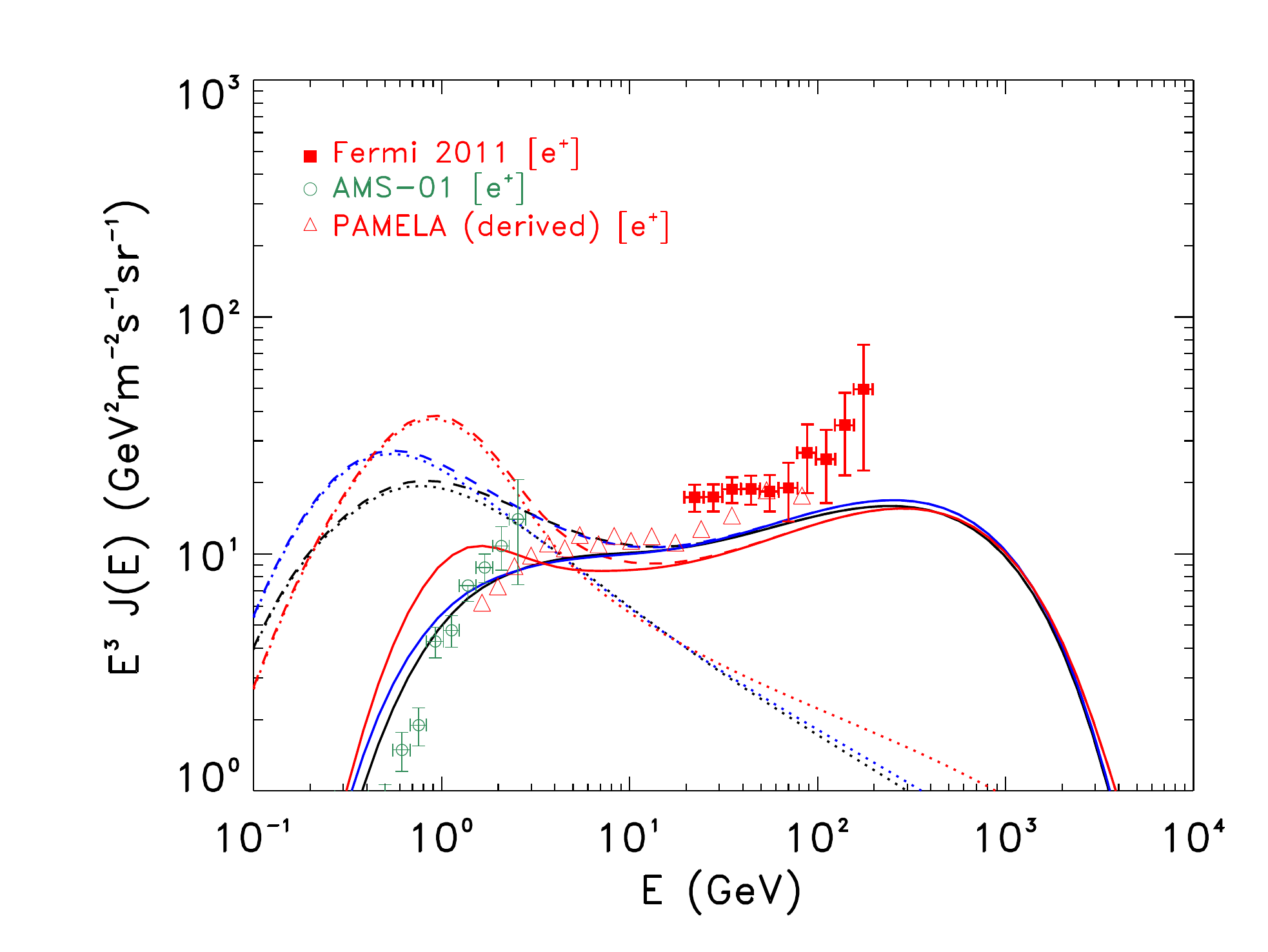}
  \caption{The $e^+$ spectrum. The line notation is the same as for Fig.\ref{fig:electrons}. PAMELA data have been derived (without error propagation) from the positron fraction and the $e^-$ spectrum released by the same collaboration. }
  \label{fig:positrons}
 \end{figure}
 
 \begin{figure}[t]
  \centering
  \includegraphics[width=0.5\textwidth]{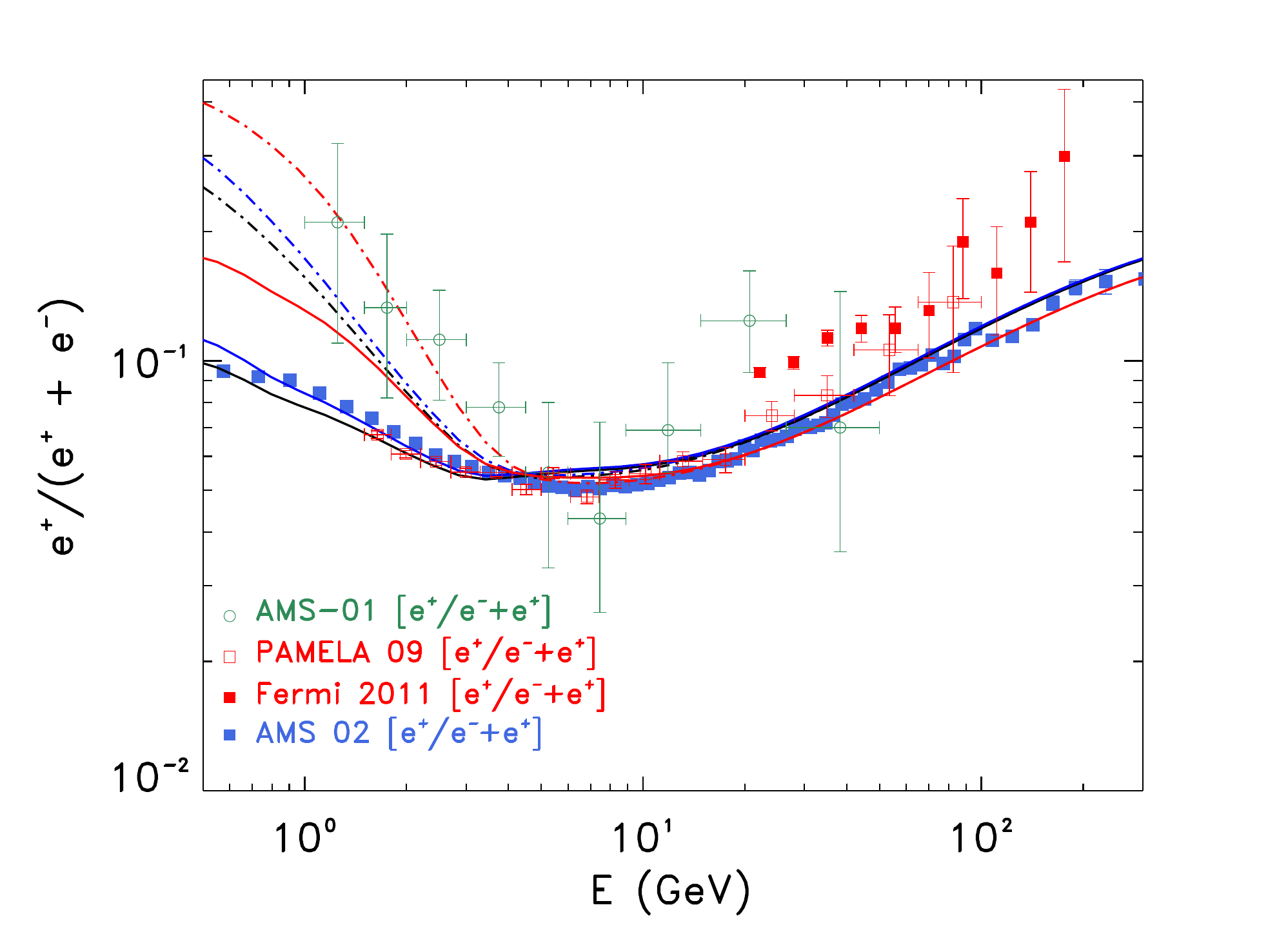}
  \caption{The PF is compared with experimental data. The color notation is the same as for Fig.\ref{fig:electrons}. Solid (dot-dashed) lines refer to AMS-02 (AMS-01) data taking periods.}
  \label{fig:posfrac}
 \end{figure}

\section{The vertical distribution of CR electrons}

To determine the vertical (perpendicular to the Galactic plane) extension of the CR diffusion region is one of the main goals in modern astroparticle physics. 
This quantity is crucial not only for conventional CR physics but also for dark matter (DM) indirect search since the local flux of DM decay/annihilation products is expected to depend significantly on it (see \cite{Evoli:2011id} for a recent analysis).  
Generally this quantity is constrained on the basis of CR radionuclides, on the $^{10}$Be/$^9$Be ratio most commonly.  This approach, however, is seriously affected by the  uncertainties related to the local distribution of sources/gas, and by solar modulation.  

The synchrotron emission of the Galaxy offers a much more direct probe of the scale height $z_t$.
First of all we notice that, when a realistic vertical distribution is adopted for the radiation interstellar field and for the GMF, energy losses in GeV energy do not affect significantly the CRE vertical distribution which is determined by diffusion and therefore coincides with that of CR nuclei (see Fig. 5 in \cite{DiBernardo:2012zu}). 

\begin{figure}[ht]
  \centering
  \includegraphics[width=0.5\textwidth]{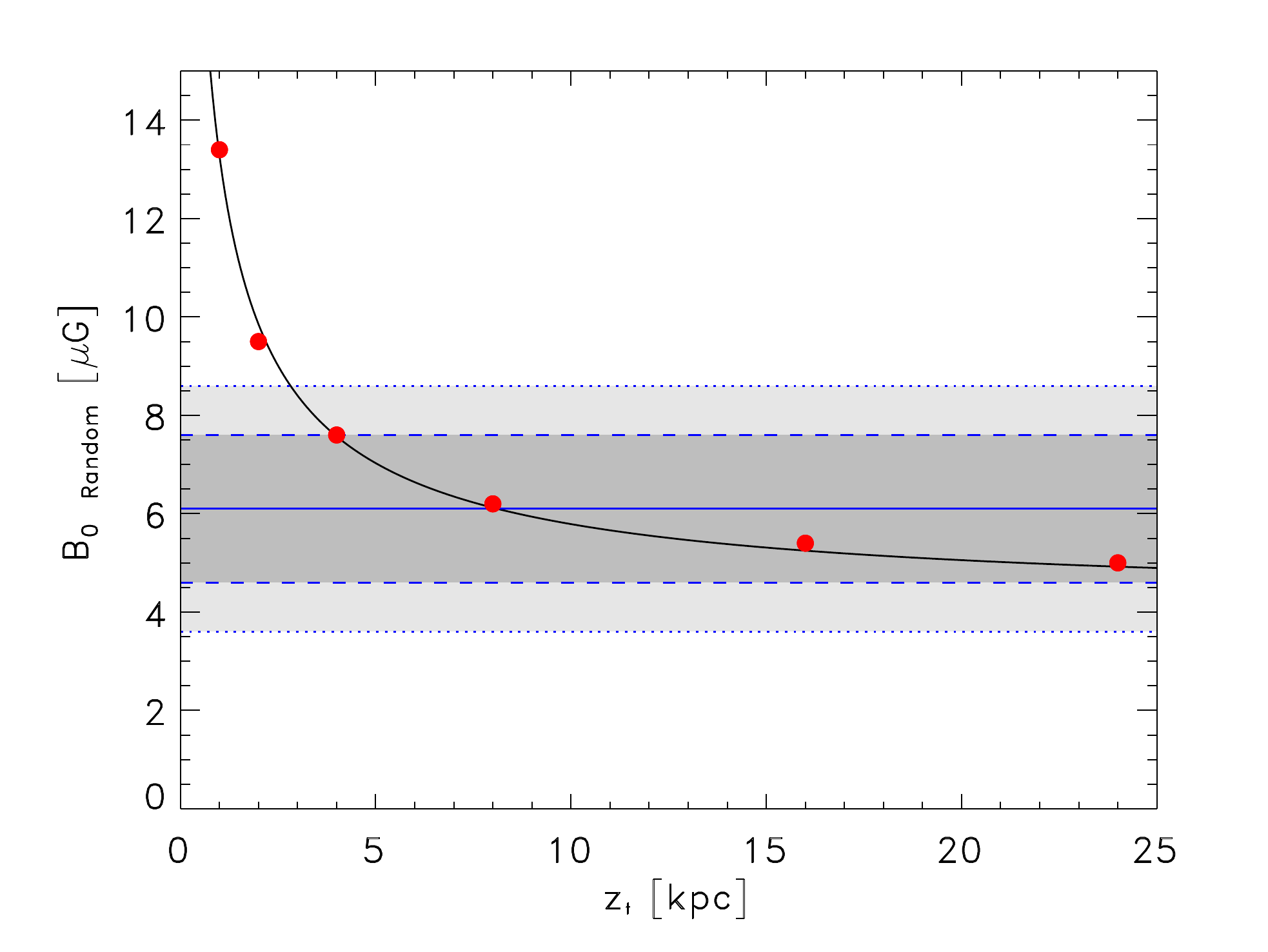}
  \caption{The normalization of the random GMF  is plotted against its vertical scale height (an exponential vertical profile is adopted for this figure).
The $3(5)~\sigma$ regions allowed by RM data are represented in gray (light-gray). Red dots are our results obtained under the condition that KRA models reproduce the observed synchrotron spectrum.  The black line is a $B_{\rm ran}^2 \propto 1/z_t$ fit of those points. The fits computed for the other setups considered in this work superimpose to that line.}
  \label{fig:zt_constr}
 \end{figure}

\begin{figure}[ht]
  \centering
  \includegraphics[width=0.5\textwidth]{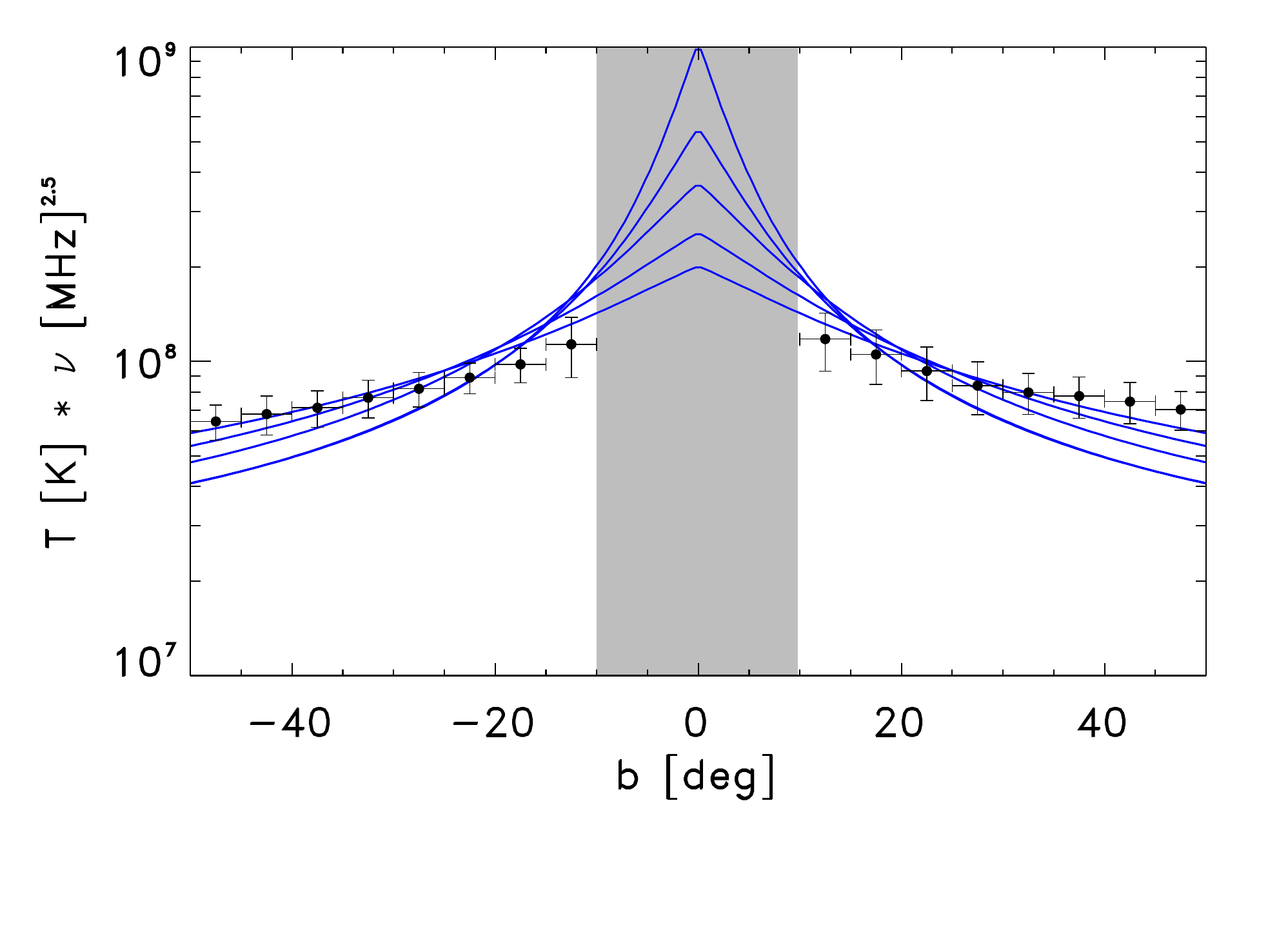}
  \caption{The latitude profiles of the synchrotron emission at 408 MHz in the region 
$40^\circ < l < 100^\circ$ computed for the KRA propagation setup and $z_t = 1,2,4,8,16~\kpc$ are compared with radio data. The grey shadowed region is not considered when placing the constraint.}
  \label{fig:lat_prof}
 \end{figure}

We consider two main independent methods to constrain $z_t$:
1) In \cite{DiBernardo:2012zu} we used \dragon to show as radio data imply a tight relation $B^{\rm rms}_{\rm ran}(z = 0) \propto z_t^{-1}$.
Then we exploit this relation to constrain $z_t$ under the condition that the random GMF fulfills the observational result $B^{\rm rms}_{\rm ran}(z = 0) = 6.1 \pm 0.5~\muG$ \cite{Han:2004aa}. This turns into the constraints $z_t >  4(3)~\kpc$ at $3 (5)\sigma$ (see Fig. \ref{fig:zt_constr}) for an exponential vertical profile of the random MF  and  $z_t >  3(2)~\kpc$ at $3 (5)\sigma$ for a Gaussian profile. In the latter case we also get the upper limit $z_t < 9~\kpc$ at $3 \sigma$. 
2) We also constrain our models against the vertical profile of the synchrotron emission latitude profile (see Fig.\ref{fig:lat_prof}).  This is similar to what done in \cite{Bringmann:2011py} where, however, only an unrealistic step-like profile was adopted for the magnetic field and diffusion coefficient.  
Furthermore, we performed this comparison in a different sky region ($100^\circ < b < 40^\circ$) where the effects of local radio sources is minimal.  
We found in this way that values $z_t \lsim 2~\kpc$ are excluded at 99 \% C.L. 
We notice that the previous analysis were performed for KRA models whose parameters, for each value of $z_t$, were tuned to minimize the combined $\chi^2$ respect to the B/C and proton data.  Different propagation models allowed by experimental data would result in very similar constraints. 

\section{Conclusions}

We showed that diffuse radio emission of the Galaxy can be used to determine the low energy tail of the interstellar electrum spectrum.   
While above few GeV this is consistent with the direct measurements performed by AMS-01, PAMELA and Fermi-LAT, below that energy, where those results are affected by solar modulation, a pronounced spectral hardening is required. 
As a consequence the electron and positron spectra below that energy are dominated by secondary particles.
We then showed that using a proper treatment of solar modulation, which accounts for the complex and time dependent structure of the magnetic fields in the heliosphere,  
the time dependent PF measured by CAPRICE, AMS-01, PAMELA and AMS-02 can consistently be reproduced. 
PF data above 10 GeV require to introduce a new electron of positron primary component with a spectral index $\sim - 1.75$.  This is lower than determined on the basis of pre-AMS-02 results.
The absolute positron spectrum measured by some of those experiments is also correctly described though only by low reacceleration models. 
This conclusion is in agreement with \cite{Strong:2011wd} and it strengthens the results of our previous analyses based on antiprotons \cite{DiBernardo:2009ku}.  

We also investigated the vertical extension of the CRE distribution by means of two independent methods.
Our results imply a lower bound on the scale height of this distribution $z_t > 2~\kpc$ and favor even higher values. 
We notice that our constraints on $z_t$ have relevant implication for the constraints on DM annihilation cross section based on the CR antiproton spectrum.

\vspace*{0.5cm}

\end{document}